\documentclass[useAMS,usenatbib]{mn2e}
\usepackage{graphicx}
\def\arcsec{$^{\prime\prime}$}

\usepackage{latexsym,amsmath,amssymb}
\usepackage{natbib}
\usepackage{rotating}
\bibpunct{(}{)}{;}{a}{}{,}

\def \nh {$N{\rm _H}$}

\def \hcm {\hbox {\ifmmode $ atom cm$^{-2}\else atom cm$^{-2}$\fi}}

\def \arcsec {\hbox{$^{\prime\prime}$}}

\def \chisq {$\chi ^{2}$}
\def \rchisq {$\chi_{\nu} ^{2}$}
\def\approxgt{\mathrel{\hbox{\rlap{\lower.55ex \hbox {$\sim$}}
        \kern-.3em \raise.4ex \hbox{$>$}}}}
\def\approxlt{\mathrel{\hbox{\rlap{\lower.55ex \hbox {$\sim$}}
        \kern-.3em \raise.4ex \hbox{$<$}}}}

\def \XMM {$XMM$-$Newton$\ }
\def \HST  {$HST$}

\def \nh {$N{\rm _H}$}

\title[Detection of Extended Emission from T Pyx]
{The  \emph{XMM-Newton}  Detection of Extended Emission from the
       Nova Remnant of T Pyxidis} 
\author[\c{S}. Balman]{\c{S}. Balman$^{1, 2}$\thanks{E-mail:\,solen@astroa.physics.metu.edu.tr}
\\
$^{1}$Department of Physics, Middle East Technical University,
In\"on\"u Bulvar{\i}, Ankara, Turkey\\
$^{2}$Department of Physics, University of Warwick, Coventry CV4 7AL, UK
}
\begin{document}

%\pagerange{\pageref{firstpage}--\pageref{lastpage}} \pubyear{2009}

\maketitle

%\label{firstpage}

\begin{abstract}

We report the detection of an extended X-ray nebulosity 
with an elongation from northeast to southwest in excess of 15$^{\prime\prime}$
in a radial profile and imaging of the recurrent nova T Pyx using the archival 
data obtained with
the X-ray Multi-Mirror Mission (\XMM), European Photon Imaging Camera 
(pn instrument). 
The signal to noise ratio (S/N) in the extended emission (above the
point source and the background) is 5.2
over the 0.3-9.0 keV energy range and 4.9 over the 0.3-1.5 keV energy range.
We calculate an absorbed X-ray flux  of 
2.3$\times$10$^{-14}$ erg cm$^{-2}$ s$^{-1}$ with a
luminosity of 6.0$\times$10$^{32}$ erg s$^{-1}$ from the 
remnant nova in the 0.3-10.0 keV band.
The source spectrum is not physically consistent with a blackbody
emission model as a single model or a part of a two-component model fitted to 
the \XMM data
({$kT{\rm _{BB}}$} $>$ 1 keV).
The spectrum is best described by two MEKAL plasma emission models 
with temperatures at 
0.2$^{+0.7}_{-0.1}$ keV and 1.3$^{+1.0}_{-0.4}$  keV. The
neutral hydrogen column density derived from the fits is
significantly more in the hotter X-ray component than the cooler one
which we may be attributed to the elemental enhancement of nitrogen and oxygen
in the cold material within the remnant.
The shock speed calculated from the softer X-ray component of the spectrum
is 300-800 km s$^{-1}$ and is consistent with the expansion speeds of the 
nova remnant derived from
the $Hubble\ Space\ Telescope$ (\HST) and ground-based optical wavelength data.
%The extened emission in the soft X-ray energy band and the soft X-ray spectrum
%are  consistent with the  optical expansion speeds of the nova shell
%found from Hubble Space Telescope (HST) data (Shara et al. 1997).
Our results suggest that the detected X-ray emission may be
dominated by shock-heated gas from the nova remnant.
%Since the emission from the central source is contraversial
%in the X-rays (see Selvelli et al. 2008) together with the fact that the
%radial profile shows significant deviations from a point source distribution
%suggest that the detected X-ray emission is dominated by shock-heated gas from the nova remnant.

\end{abstract}


\section{Introduction}

Classical novae (CNe) outbursts are the explosive ignition of
accreted material on the surface of the white dwarf (WD) in a
cataclysmic variable (CV) as a result of a thermonuclear
runaway causing
the ejection of 10$^{-7}$ to 10$^{-3}$ M$_{\odot}$ of material at velocities
up to several thousand kilometers per second (Shara 1989; Livio 1994; Starrfield 2001; Bode $\&$ Evans 2008).
Though there has only been one previous (and one very marginal)
detection of old CNe remnants
in the X-ray wavelengths (Balman \& \"Ogelman 1999; Balman 2005,2006; 
Pek\"on \&  Balman 2008),
CNe have been detected in the hard X-rays (above 1 keV) as a
result of accretion, wind-wind and/or blast wave interaction
{\it in the outburst stage}
(O'Brien et al. 1994; Krautter et al. 1996; Balman, Krautter, \"Ogelman 1998, Mukai \& Ishida 2001, Orio et al. 2001; Ness et al. 2003;
Hernanz \&  Sala 2002,2007; Page et al. 2009).
Recurrent novae (RNe) are a type of CNe where outbursts recur
with intervals of several decades (Webbink et al. 1987; Hachisu $\&$ Kato 2001; Bode \& Evans 2008).
In general, these systems are expected to have high
accretion rates of  10$^{-8}$ to 10$^{-7}$ M$_{\odot}$ yr$^{-1}$  onto 
massive WDs close to the Chandrasekhar limit;
occurrence of recurrent outbursts in relatively less massive WDs is also 
possible
(Starrfield, Sparks, Truran 1985; Prialnik \& Kovetz 1995; Yaron et al. 2005).
RNe are detected in the hard X-rays
{\it in the outburst stage} as a result of  wind-wind and/or 
blast wave interaction
during the outburst stage (Orio et al. 2005; Greiner  \& Di Stefano 2002; Bode et al. 2006;
Sokoloski et al. 2006; Drake et al. 2009; Ness et al. 2009). 
Recently, extended X-ray emission 
associated with the radio jet of the recurrent nova RS Oph
is discovered (Luna et al. 2009).

Recurrent nova T Pyx has 5 recorded outbursts in
1890, 1902, 1920, 1944, and 1966 
(Webbink et al. 1987). Ground-based CCD observations (Shara et al. 1989) show 
the existence of at least two of the shells extending to a size
of $\sim$ 20$^{\prime\prime}$ in diameter and also a faint [OIII] 
shell has been found.
$Hubble\ Space\ Telescope$ (\HST; 1994-2007) and ground-based imaging of the shell 
around T Pyx show thousands of knots
in H$\alpha$ and [NII] with expansion velocities of about 350-715 km
s$^{-1}$, with shell expansion speed around 500  km s$^{-1}$
(Shara et al. 1997; O'Brien \& Cohen 1998; Schaefer, Pagnotta \& Shara 2010).
The \HST\ observations support an interacting shells model producing the clumping, 
shock heating, and emission lines. Schaefer et al. (2010) show that most of the
knots have not decelerated and are powered by the RN outbursts and originate from a
CN outburst of the year 1866. 
T Pyx is suggested to be a wind-driven source
(due to the high mass accretion rate of $\dot M$ $\sim$ 1$\times$
$10^{-7}$ M$_{\odot}$ yr$^{-1}$) and a Super Soft X-ray source (SSS) 
(Patterson et al. 1998). On the other hand, 
Greiner  \& Di Stefano (2002) reports a 
ROSAT non-detection of the source in December 1998 excluding the possibility 
of the existence of a SSS.
Gilmozzi \& Selvelli (2007) and Selvelli et al. (2008)
show that the UV+opt+IR spectrum of T Pyx is dominated by
the accretion disk and the continuum in the UV can be represented by a
blackbody of T $\sim$ 34000 K with $\dot M$ $\sim$ 1$\times$10$^{-8}$ M$_{\odot}$ yr$^{-1}$.
Their detailed study based on the UV data excludes the possibility
that T Pyx is a SSS and Selvelli et al. (2008) uses 
this X-ray Multi-Mirror Mission (\XMM) 
data set to show that a SSS nature is not supported.
%Schaefer et al. (2010) suggest that 
%the decline in the observed accretion rate does not sustain a SSS 
%and T Pyx has an evolutionary cycle
%going from a normal CV state to its present RN state and to a future
%hibernation state.

\section{The Data and Observation}

The \XMM Observatory (Jansen et al. 2001) has three 1500 cm$^2$ 
X-ray telescopes each with an European Photon Imaging Camera (EPIC) 
at the focus; two of which have Multi-Object Spectrometer (MOS) CCDs
(Turner et al. 2001) and the last one uses pn CCDs (Str\"uder et al. 2001) 
for data recording.
%Also, there are two Reflection Grating Spectrometers 
%(RGS, den Herder et al. 2001).
T Pyx was observed (pointed observation) by \XMM for a duration of 51 ks 
between
2006 November 9 UT 19:22:59 and 2006 November 10 UT 09:02:31 with a 
slight off-axis angle of about 1$^{\prime}$.
A medium optical blocking filter was used with all the EPIC cameras and the 
pn, MOS1 and MOS2, instruments were operated  in the full frame imaging mode. We
analysed the pipeline-processed data using Science Analysis Software (SAS) 
version 8.0.5. Data 
(single- and double-pixel events, i.e., patterns 0--4 with Flag=0 option) were extracted from
a circular region of radius 15$^{\prime\prime}$ for pn, MOS1 and MOS2
in order to perform spectral analysis together with the background events extracted from a
source free zone normalised to the source extraction area. In this paper, 
we will
present an image obtained by the EPIC pn since it has more source 
photons
due to its better sensitivity compared with the MOS1 and MOS2 instruments. However, 
we simultaneously
use EPIC pn, EPIC MOS1 and EPIC MOS2 data to determine the X-ray spectrum of the source and increase the
number of data points in the fitting process.
We cleaned the pipeline-processed event file from  the existing flaring 
episodes by creating user-select
good time intervals (gti) with a count rate threshold 
$<$ 0.08 c s$^{-1}$ for the two MOS and $<$ 0.11 c s$^{-1}$ for the pn
instruments over the 0.3-9.0 keV energy range.
This method cleaned the flares.
Within the extraction area indicated above, the final source count rates were
0.007$\pm$0.001 c s$^{-1}$ for pn,
0.003$\pm$0.001 c s$^{-1}$ for MOS1 and MOS2 with the effective
exposure times of 30.6 ksec, 39.7 ksec and 38.8 ksec for the pn, MOS1 and MOS2 instruments, 
respectively. 
%We double-checked the cleaning process with the removal of 
%all times once the flaring started and this method yields similar count rates
%for the three detectors.

\section{Analysis and Results}

\subsection{The X-ray Image and the Radial Profile of T Pyx}
\label{subsect:ima}

The pipeline-processed and cleaned event files are used to calculate
the X-ray image of T Pyx and its vicinity. The largest size of the shell
of T Pyx is determined to be 10$^{\prime\prime}$ in radius (see the Introduction). 
Figure 1 shows an  X-ray intensity image between 0.3 and 9.0 keV obtained from 
the EPIC pn data.
The EPIC pn pixels are binned by 20 pixels (each 0$^{\prime\prime}$.05) and the PI channels are filtered between 0.3-9.0
keV in order to create an image with a pixel resolution of 1$^{\prime\prime}$ in the sky.
The image is then smoothed by a (variable) Gaussian of $\sigma$=1$^{\prime\prime}$-2$^{\prime\prime}$ 
using a minimum significance
of 1 and a maximum significance of 3 $\sigma$. Since adequate filtering on the 
background events are used (see section 2), 
the background has not been subtracted out from the image.
The small red circle
indicates the position of T Pyx. Additionally, in Figure 1, the X-ray intensity
contours are overlaid using
linear increments of intensity. The image shows asymmetry
in the emission around the central source position
with a minimum elongation from the northeast to the southwest 
of  about 15$^{\prime\prime}$ in size
(as opposed to the width of the source as detected 
in the image of about 5$^{\prime\prime}$ only).
%The width of the
%source image is less than 5$^{\prime\prime}$ at all locations. 
%There seems to be an asymmetry of very 
%faint emission on the western side of the point source as opposed to east
%within 10$^{\prime\prime}$ radius. 
%The red circle showing the point source
%location is slightly off-set (about 2 arcsec) from the peak of emission as can be seen in the X-ray
%contours.

We also constructed radial profiles in the 0.3-9.0 keV,
0.3-1.5 keV and 2.0-9.0 keV energy bands from images created
 at 1$^{\prime\prime}$ pixel resolution (unsmoothed images) 
using the SAS task {\tt ERADIAL} keeping the centroid fixed at the
source position (J2000) of T Pyx.
{\tt ERADIAL} is a routine to extract a radial profile of a source in an
image field and fit a point spread function (PSF) to it. Figure 2 shows the 
radial profiles
and the normalised EPIC pn PSF (to the source counts in the 
radial bins 2$^{\prime\prime}$-4$^{\prime\prime}$ 
from the position of T Pyx) in 
three different energy 
bands. The solid line is the normalised PSF
and the data are indicated by unfilled squares and vertical error bars.
A fitted background (using radial profiles)  is also added to the PSFs 
during the normalisation process.
%Figure 2 shows that radial distribution of the
%counts (azimuthally averaged) in the vicinity of T Pyx does not follow the 
%PSF of the EPIC pn instrument.
The top and middle panels of Figure 2 show significant deviation between the
PSF and the radial profile from 6$^{\prime\prime}$ out to 
10$^{\prime\prime}$-30$^{\prime\prime}$ from 
the location of the central binary system, and 
the bottom panel shows  marginal variations.
The EPIC pn PSF is described in Ghizzardi (2002) with a King profile (model) 
whose core radius and power-law index are calculated to be about
6$^{\prime\prime}$-4$^{\prime\prime}$ and  -(1.5-1.4), respectively, 
for off-axis angles $\le$ 1$^{\prime}$
over the entire energy band of \XMM . 

A King profile is described by $A[1+(r/r_0)^2]^{- \alpha}$ where
$r_0$ is the core radius and  $\alpha$ is the power-law index. 
%The fit using a
%King model to the radial profile in the $top\ panel$ of Figure 2 yields a 
%\rchisq\ value in excess of 10.0
%which shows that it is not a King model/profile. 
We fitted the radial profiles (in Figure 2) with a composite model of a King profile
and a constant for the background. The fit to the radial profile in the top panel 
yields a \rchisq\ value of 1.3 with 
the best fitting core radius  in a range
0$^{\prime\prime}$.2-1$^{\prime\prime}$.0 and the power-law index  between 
0.4 and 0.7 (ranges
correspond to errors at 99\% confidence level). Freezing
the two parameters to the acceptable values of the EPIC pn PSF (
6$^{\prime\prime}$ and -1.5) changes the  \rchisq\ value to 5.5 .
The fit to the radial profile in the soft energies (
the $middle\ panel$ of Figure 2) yields a core radius 
in a range 1$^{\prime\prime}$.9-4$^{\prime\prime}$.0 and the power-law index is between 0.5 and 0.8
(ranges correspond to errors at 99\% confidence level). \rchisq\ of this fit is 1.0 . 
This is also not consistent with the King profile parameters of the EPIC pn 
PSF (at 3$\sigma$ confidence level). Fixing the 
core radius at 6$^{\prime\prime}$ and the power law index to -1.5
results in a \rchisq\ of 2.4. 
%Keeping the power-law index frozen at a value of 1.5 yields
%a core radius between 10$^{\prime\prime}$ and 46$^{\prime\prime}$ 
%with a \rchisq\ of 2.5 . 
None of the fits to the radial profile of the soft band yield parameters consistent with
the EPIC pn PSF at 3 $\sigma$ confidence level.
The fit to the profile in the bottom panel (hard energies) 
yields a \rchisq\ lower than 2 when parameters are fixed at the expected values of
the EPIC pn PSF.
 
Using the radial profiles,
we calculate a signal-to-noise ratio 
of 5.2 in the 0.3-9.0 keV and a ratio of 4.9 in the 0.3-1.5 keV range for the extended emission from
T Pyx (Signal-to-noise ratio is S/$\sqrt{B+P}$; S is the net extended source counts, B is the total background 
counts and P is the counts within the PSF in an annular region of inner radius 6\arcsec\ and outer radius 16\arcsec\ ).  
We, also, searched the \XMM archival database for pointed observations of CVs and checked how the 
PSF normalised
with the radial profiles using the same method and found that GK Per shows a similar discrepancy from the 
point source distribution as
would be expected since GK Per is an extended X-ray source. Furthermore, we checked the radial profiles
of other closeby weak sources in the field of T Pyx and found that the radial profiles mostly obey the
model PSF (taking the off-axis angles into account).

\begin{figure}
\includegraphics[width=3.20in,height=2.7in]{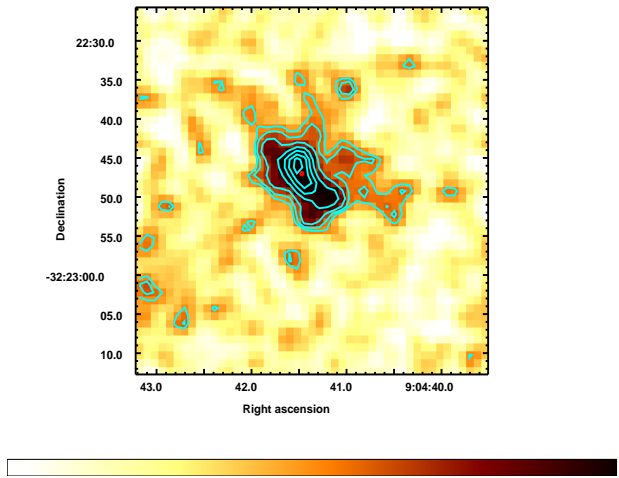}
\caption{The X-ray image of  the vicinity of T Pyx in the
0.3 to 9.0 keV range. The pixel resolution is
1$^{\prime\prime}$. North is up and West is to the right.
The small red circle shows the location
of the binary system. Overlaid are the X-ray contours with linear intensity increments.}
\end{figure}

\begin{figure}
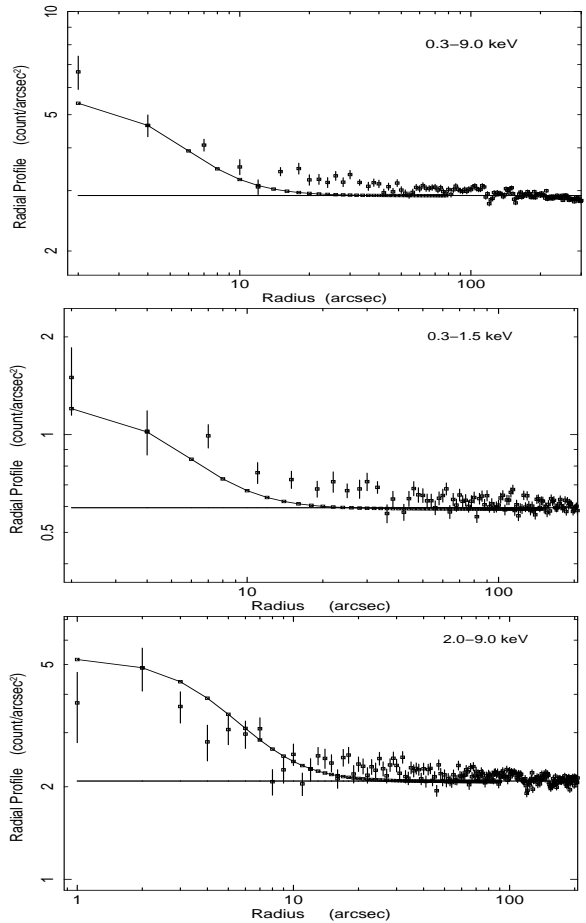

\hspace{-0.05cm}
\includegraphics[width=1.6in,height=3.0in,angle=-90]{fig2_a_son1.ps}\\
\includegraphics[width=1.6in,height=3.0in,angle=-90]{fig2_b_son1.ps}\\
\includegraphics[width=1.6in,height=3.0in,angle=-90]{fig2_c_son.ps}
\caption{The radial profiles of the source counts in the vicinity of  T Pyx
in three energy bands: 0.3-9.0 keV, top panel, 0.3-1.5 keV, middle panel, 
and 2.0-9.0 keV, bottom panel.
The flat solid lines are the background. The source
radial profiles are the data points indicated by unfilled squares and vertical
error bars. The  EPIC pn PSFs are indicated as curves with solid lines 
and are normalized to the counts in the first two radial bins of the profiles.
A background is also added to the PSFs during the normalisation process.
The resolution of spatial binning is 2$^{\prime\prime}$ for the top and middle
panels and 1$^{\prime\prime}$ for the bottom panel with a few consecuitive bins 
averaged to improve statistics.
The radial profiles and the corresponding EPIC pn PSFs are prepared 
using the SAS task {\tt ERADIAL}.
}
\end{figure}

\subsection{The \XMM Spectrum of T Pyx}
\label{subsect:spectra}

We performed spectral analysis of the EPIC data using the SAS task 
{\tt ESPECGET}
and derived the spectra of the source and the background together with the appropriate response 
matrices and
ancillary files. How the photons were extracted is described in Section 2.
The EPIC pn, EPIC MOS1 and EPIC MOS2 spectra were simultaneously fitted
to derive spectral parameters of the emission arising from 15$^{\prime\prime}$ 
(radius of the circular photon exraction region) of the source 
position. The
spectral analysis was performed
using XSPEC version 12.5.1 (Arnaud 1996). A constant factor
was included in the spectral fitting to allow for a normalisation
uncertainty between the EPIC pn and EPIC MOS instruments.
We grouped the pn and MOS spectral energy channels  in
groups of 100-150 to improve the statistical quality of the spectra.
The fits were conducted in the 0.3-9.0 keV range. 
At first, we fitted a
single blackbody or a composite model spectra where one component was a 
blackbody model to check the SSS scenario for the binary system.
We find that the fits yield 
temperatures in excess of 1 keV as the best fit value which is inconsistent with
the SSS scenario in accordance with the results of Selvelli et al. (2008).
Next, assuming the emission is of the accretion shocks or the shocks in the 
nova ejecta, 
the three spectra were fitted with a single or two MEKAL emission models 
representing thermal plasma emission in collisional equilibrium
(Liedahl, Osterheld \& Goldstein 1995).
%This model was used because it is expected that the shocked material in a 
%nova ejecta is best modeled
%as plasma in collisional equilibrium given that enough time 
%has passed since the eruption.
For the intervening absorption (\nh) we used the
TBabs multiplicative model in XSPEC (Wilms, Allen \&  McCray 2000).
The \nh\ was let to vary in order to account for
any intrinsic absorption. All abundances were assumed at their solar values.
We find that the best fits to the data using a single or two MEKAL models 
(TBabs*MEKAL or TBabs*(MEKAL+MEKAL)) 
yield unbounded plasma temperatures in excess of 70 keV with reduced $\chi^2$ values of 1.3-1.5. 
%resulting in nonphysical spectral parameters. 
%A 70 keV X-ray temperature is
%particuarly inconsistent with the extended nebulosity in the soft X-rays from 0.3-1.5 keV. 
High shock
temperatures may be achieved in accretion shocks, however
we note that T Pyx is a
non-magnetic CV and such systems show X-ray shock temperatures of 3-10 keV, in general
(Baskill, Wheatly \& Osborne 2005).
Such high temperatures ($\sim$ 70 keV) are mostly achieved by some 
intermediate polars (a subclass of magnetic CVs)
which T Pyx does not  belong to (see Brunschweiger et al. 2009).
We find that these fits are 
inconsistent with the accreting CV observations. 
Such high X-ray temperatures are also not in accordance
with the hard X-ray observations of 
CNe/RNe (see the references in the Introduction). 
%Next, we applied a two-component MEKAL model
%fit to the spectra (TBabs*(MEKAL+MEKAL)) which, also, yielded 
%the same unbounded plasma
%temperatures for both of the MEKAL
%components as above with reduced $\chi^2$ value larger than 1.5. 
Furthermore, in the fitting process, we used some other plasma
emission models like CEVMKL and MKCFLOW within the XSPEC software (instead of MEKAL model) 
that are largely used for the accreting CVs. The resulting X-ray
temperatures are, also, unbounded and above 99 keV for CEVMKL and 
21 keV-73 keV (lowT-highT) for MKCFLOW 
which are, also, inconsistent with CV observations. 
Finally, a successful fit was achieved 
using two different absorption components
together with the two different temperature MEKAL models 
(TBabs*MEKAL+TBabs*MEKAL) with a reduced $\chi^2$ of 1.0 for
 15 $d.o.f$ (degrees of freedom).
Figure 3 shows the EPIC pn , EPIC MOS1 and EPIC MOS2 spectra fitted with this 
composite model.
We derive for the first emission component an $N{\rm _{H1}}$ of  0.6$^{+0.5}_{-0.4}$$\times 10^{22}$ cm$^{-2}$, $kT{\rm _{1}}$ of
 0.2$^{+0.7}_{-0.1}$ keV, and a normalisation of 2.9$^{+168.0}_{-2.7}$$\times 10^{-4}$. 
The second emission component
has an $N{\rm _{H2}}$ of  5.5$^{+11.5}_{-4.0}$$\times 10^{22}$ cm$^{-2}$, 
$kT{\rm _{2}}$ of 1.3$^{+1.0}_{-0.4}$ keV, and a
normalisation of 4.5$^{+9.3}_{-1.7}$$\times 10^{-4}$.
Spectral uncertainties are given at 90\%
confidence level ($\Delta$\chisq = 2.71 for a single parameter).
The best fit results above indicate an absorbed X-ray
flux of 2.3$\times 10^{-14}$ erg s$^{-1}$ cm$^{-2}$ and an unabsorbed X-ray flux of 3.7$\times 10^{-13}$ erg s$^{-1}$ cm$^{-2}$ which translates to
an X-ray luminosity of 6.0$\times 10^{32}$ erg s$^{-1}$ at the source distance of 3.5 kpc 
(distance from Selvelli et al. 2008) in the energy range 0.3-10.0 keV. 
We also checked the existence of two spectral components by creating spectra in 
two annular photon extraction regions: (1) inner radius of 0\arcsec\ and outer radius of 6\arcsec, and
(2) inner radius of 7.5\arcsec\ and outer radius of 15\arcsec\ , centered on T Pyx. 
We find a similar spectrum to
Figure 3 in the inner extraction region. However, the spectrum of the outer annulus shows only a
65$\%$ decrease in the emission below 1.5 keV and no emission above 1.5 keV. There is no
need for a second harder X-ray spectral component.

\begin{figure}
\includegraphics[width=2.4in,height=3.1in,angle=-90]{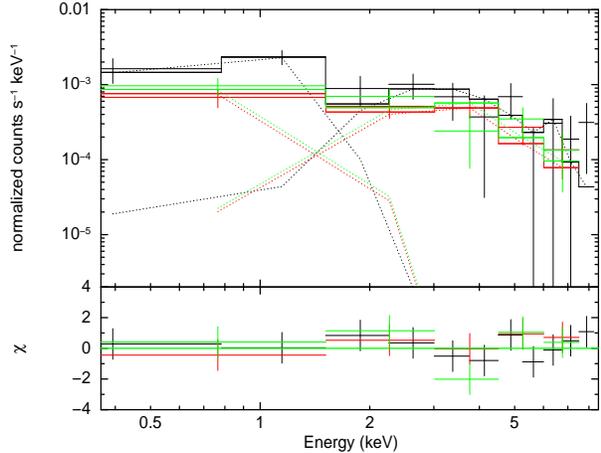}
\caption{The \XMM\ EPIC data fitted with (TBabs*MEKAL + TBabs*MEKAL) model of emission in the 0.3-9.0 keV band.
The red, green and black crosses indicate the EPIC MOS2, EPIC MOS1 and EPIC pn  data, respectively.
The dotted lines show the contribution of the two fitted MEKAL models.
The lower panel shows the residuals between the data and the
model in standard deviations. }
%\label{fig:spect}
\end{figure}

\section{Discussion}

%The circumstellar interaction models for RNe indicate the existence
%of forward and reverse
%shocked material traversing into the circumstellar medium created by the different shells
%expelled during previous
%outbursts or a preexisting red giant wind, etc.
%(O'Brien, Bode \& Kahn 1992; Contini \& Prialnik 1997; Vaytet, O'Brien \& Bode 2007; Walder,
%Folini, Shore 2008).  CNe and RNe remnants may
%resemble  Supernova remnants (SNRs) in several aspects as the ejected material  
%(in a blast wave) interacts with the circumstellar medium.
%The study of the old nova shell of GK Per (Nova Per 1901)  in the X-rays shows the 
%interaction of a blastwave with its surroundings and how it evolves
%like miniature SNR (Balman 2005; Balman \& \"Ogelman 1999). The recent outburst of the
%recurrent nova RS Oph has also shown a blast wave interaction with non-thermal
%radio emission and collimated outflows, however the remnant can not be resolved
%in the X-rays due to its young age
%(Eyres et al. 2009; Drake et al. 2009; Bode et al. 2006;
%Sokoloski et al. 2006).

The spectrum of T Pyx is well described by two different plasmas in 
ionization equilibrium at different temperatures.
We stress that the expected SSS is not detected and the radial profiles
deviate significantly from the PSF of EPIC pn. If one assumes 
all the detected luminosity is due to accretion, this yields an accretion rate 
less than $<$ a few$\times$10$^{-10}$ M$_{\odot}$ yr$^{-1}$ for the system in
contradiction with the optical and UV measurements by a factor of 10-100.
%Selvelli et al. (2008) predicts
%an X-ray luminosity of 10$^{35}$ erg\ s$^{-1}$, if the emission is 
%of the shocks in the accretion disk, which is about a factor of 
%50-100 more than what we detect using the \XMM data.
This strengthens the possibility that most of the detected emission
is of the nova remnant in the X-rays rather than the point source.
%However, we caution that one of the components may still belong to the
%binary system. 
%Given these caveats,
%we interpret the derived spectral parameters in terms of shock-heated shells.

A simple shocked-shell model is of thermal origin.
The total power from the shocked-shell, as an X-ray emitting nebula,
can be expressed as in Balman (2005, 2006)
$ L_x\simeq 3.1\times 10^{33} T_{7}^{0.5} n_o^2  R_{18.5}^{3}$.
The temperature $T$ is in units of 10$^7$ K and radius of the shell
$R$ is in units of 3.1$\times 10^{18}$ cm. Using
$R\sim4.2\times 10^{17}$ cm, $n_o\sim$1-50 cm$^{-3}$
and $T\sim 10^{6.3-7.3}$ K,  $L_{\rm x}$ is about
1.0$\times 10^{31}$-8.0$\times 10^{33}$ ergs s$^{-1}$. The detected X-ray
luminosity (with \XMM) is in this expected range of $L_{\rm x}$ for 
the shocked-shell emission.
The detected emission measure $EM = <n_e>^2 V_{eff}$ (calculated from normalisation of the fit)
yields an average electron density $n_e$ of about 4.7 cm$^{-3}$ and 5.5 cm$^{-3}$
for the colder and hotter plasma, respectively,
using a volume of 3.0$\times 10^{53}$ cm$^{3}$ (consistent
with a spherical region of 8$^{\prime\prime}$ radius at 3.5 kpc) and  a filling factor f=1.
If the filling factor is as low as
1$\times 10^{-4}$, then the electron density can be as high as 
470-550 cm$^{-3}$ in the X-ray emitting region.
The electron density calculated for the nova shell of GK Per
is of similar order
0.6-11.0 cm$^{-3}$ for f=1 (Balman 2005).
The spectrum of the nova remnant of GK Per also shows two plasma emission 
components with a significant difference of
absorption between the two components (Balman 2005). 
Such differences can be explained
by enhanced non-solar abundances of metals in cold material 
(possibly a shell or collection of
dense knots) between the two emission components.
A simple fit with the VARABS or VPHABS model (within XSPEC) 
using variable abundances 
for the second harder X-ray component yields an  $N{\rm _{H2}}$ that is 
similar to  $N{\rm _{H1}}$ within error
ranges with enhanced abundances of N/N$_{\odot}$$\sim$ 20 
and O/O$_{\odot}$$\sim$ 10 . The remnant of T Pyx shows strong [NII] emission, but
the abundances of metals is unknown. The $N{\rm _{H}}$ difference
can also be attributed to some complicated warm absorber effect (e.g. on 
the central binary system or the inner shell).
Contini \& Prialnik (1997) have modeled the circumstellar interaction of T Pyx shells. They
find that as the latest shell catches the older shell, a forward shock moves into the older
ejecta and a reverse shock moves into the new ejecta with typical 
electron densities of 
100-300 cm$^{-3}$ and electron temperatures of about 0.1-0.9 keV. This model is very consistent 
with the X-ray spectral parameters obtained from the EPIC pn data
(soft component). 

The plasma temperatures of the two different MEKAL model components can be used to 
calculate the shock velocities using the general relation
$kT_s=(3/16)\mu m_H (v_s)^2$, ($T = 1.4\times10^5 v^2_{\ 100 km s^{-1}}$),
assuming Rankine-Huguenot jump conditions in the absence of particle acceleration. We derive
400 km s$^{-1}$ (300-800 km s$^{-1}$ using error ranges)
for the first plasma emission component and 1050 km s$^{-1}$ for the
second more absorbed plasma emission component (maximum limit 1400 km s$^{-1}$ using error ranges).
The shell expansion velocity is measured to be
about 350-715 km s$^{-1}$ 
(Shara et al. 1989; Shara et al. 1997;O'Brien \& Cohen 1998; Schaefer et al. 2010). 
This is consistent with the shock speeds calculated using the spectral parameters
of the softer X-ray component. The radial profiles calculated from the \HST\ 
data require 
a multiple shell model with a particular shell found around 5\arcsec\-6\arcsec\
and an extended emission region that goes out to 10\arcsec (both measured 
from the position of T Pyx; Shara et al. 1997; Schaefer et al. 2010). 
The [NII]+H${\alpha}$ images 
evidently shows an elongation/extension from  the northeast to the southwest
of the source position (see Figure 2c in Shara et al. 1989).
%Furthermore, 
%there is also a faint extended 
%emission region that goes out to a radius of
%10\arcsec. We make a note that the brightest point in the total 
%X-ray image is off-set from the binary system position
%by about 2\arcsec\ consistent with the \HST\ data. This is relevant only if no X-ray point source 
%is detected.
The X-ray  image indicates  a similar elongation
from the northeast to the southwest as well. 
This may be a result of the interaction between 
a bipolar outflow (e.g., a fast wind) during a RN outburst and the circumstellar
environment (e.g., different shells) of the nova or  
the suggested CN remnant of 1866 by Schaefer et al. (2010). 
%We note that the expansion speeds of this bipolar outflow
%must be inexcess of the fastest ejecta detected from the system.
%The [NII]+H${\alpha}$ images (\HST)
%evidently shows an elongation/extension from  northeast to southwest
%from the source position (see Figure 2c in Shara et al. 1989).
%Moreover, ground-based optical spectra show the existence of
%blue and red shifted [NII] emission lines ($\sim$500 km s$^{-1}$) and complex
%velocity fields in the nova remnant (O'Brien \& Cohen 1998; Bruce \& Deutsch 1998).
The expansion speeds of the 1966 outburst are in a range
850-2000 km s$^{-1}$ (Catchpole 1969).
It has been about 40 years since the last outburst and the more absorbed 
(i.e. embeded) plasma emission component may belong to the most recent outburst. 
The expected location of the ejecta is about 2\arcsec\ - 4\arcsec\ (from the
position of T Pyx). 
%However, the extension
%in the radial profile of the hard X-ray component is not well supported. 
The expected size of the new
interaction zone is within the core size of the EPIC pn PSF.  
The harder X-ray component may, also, belong to the binary system.
A long observation of this source 
using the $Chandra$ Observatory (yielding better statistics) can resolve this issue with
the  aid of its superb pixel and PSF resolution. 

%Contini \& Prialnik (1997) have modeled the circumstellar interaction of T Pyx shells. They
%find that as the latest shell catches the older shell, a forward shock moves into the older
%ejecta and a reverse shock moves into the new ejecta. 
%Typical densities they find are about
%100-300 cm$^{-3}$ and electron temperatures of about 0.1-0.9 keV are achieved. They also
%calculate a few hundred km s$^{-1}$ of shock velocity. This model is very consistent with
%the X-ray spectral parameters obtained from the EPIC-PN data 
%(soft component). We note
%that we detect  more neutral hydrogen column density associated with
%the hotter plasma component which may indicate a cold zone
%between the  interacting shells; also predicted in Contini \& Prialnik (1997).
%This could be due to non-solar abundances of elements in the cold shell  
%particularly of nitrogen, oxygen, 
%and neon (as in GK Per; Balman 2005) or
%some complicated warm absorber effect.

\section*{Acknowledgments}
The author thanks the SWIFT-NOVA group 
(http://www.swift.ac.uk/nova-cv/) for constructive comments.
SB acknowledges support
from T\"UB\.ITAK, The Scientific and Technological Research Council
of Turkey,  through project 108T735 and also support from  EUFP6 Project
MTKD-CT-2006-042722. 
SB also thanks Tom Marsh and Danny Steeghs for the hospitality
during her visit at the University of Warwick.

%\label{lastpage}

\end{document}